\documentclass[%
 aip,
 amsmath,amssymb,
 reprint,%
]{revtex4-1}

\usepackage{graphicx}
\usepackage{dcolumn}
\usepackage{bm}

\usepackage[utf8]{inputenc}
\usepackage[T1]{fontenc}
\usepackage{mathptmx}
\usepackage{etoolbox}
\usepackage{comment}
\usepackage{amsmath}
\usepackage{setspace}
\usepackage{array}
\usepackage{booktabs}
\usepackage{amsbsy} 
\usepackage{amsthm} 
\usepackage{mathptmx}   
\usepackage{float}

\usepackage[colorlinks=true,linkcolor=blue,citecolor=blue,allcolors=blue]{hyperref}

\makeatletter
\def\@email#1#2{%
 \endgroup
 \patchcmd{\titleblock@produce}
  {\frontmatter@RRAPformat}
  {\frontmatter@RRAPformat{\produce@RRAP{*#1\href{mailto:#2}{#2}}}\frontmatter@RRAPformat}
  {}{}
}%
\makeatother
\begin{document}
\preprint{AIP/123-QED}

\title
{Suspending droplets beyond the Rayleigh limit: The interplay of acoustic and gravity forces}
\author{Jeyapradhap Thirisangu}
 \affiliation{ Department of Mechanical Engineering, Indian Institute of Information Technology, Design and Manufacturing, Kancheepuram, Chennai - 600127, India}
 \author{E Hemachandran}
 \affiliation{ Department of Mechanical Engineering, Indian Institute of Information Technology, Design and Manufacturing, Kurnool, Kurnool - 518007, India}
\author{Karthick Subramani}%
 \email{karthick@iiitdm.ac.in}
\affiliation{ 
Department of Mechanical Engineering, Indian Institute of Information Technology, Design and Manufacturing, Kancheepuram, Chennai - 600127, India
}%

\date{\today}

\begin{abstract}

In this work, we experimentally investigate the suspension behavior of droplets subjected to standing acoustic waves. We focus on the droplet sizes beyond the Rayleigh limit, i.e., when the droplet size is comparable to the wavelength of the acoustic wave. We show that an acoustic field can disrupt the uniform motion of aqueous droplets in oil and cause them to either suspend or settle, depending on the interplay between acoustic and gravity forces. Remarkably, in contrast to droplets within the Rayleigh limit, the critical acoustic power or minimum pressure amplitude required to suspend droplets beyond the Rayleigh limit is dependent on droplet size. As the droplet size increases, the critical acosutic power increases significantly. Building upon this understanding, a novel sorting method is proposed based on critical acoustic power.
\end{abstract}

\maketitle

\section{{\textbf{INTRODUCTION}}}

The seminal concept of manipulating matter using acoustic waves was first demonstrated by Kundt et al.\cite{kundt1874longitudinal} during the later part of the 19th century. To explain the above phenomenon, in 1934, King \cite{king1934acoustic} conducted a detailed theoretical study and provided the expression of the radiation pressure exerted on a rigid sphere without considering the compressibility. Subsequently in 1955, Yosioka and Kawasima \cite{yosioka1955acoustic} expanded the King's theoretical framework by incorporating compressible spheres and achieved good agreement with the experimental observations of air bubbles in the water. Gor'kov\cite{gor1962forces} developed an elegant approach by formulating the acoustic radiation force as the gradient of a potential (now commonly referred to as the Gor'kov Potential) to replicate the results of Yosioka and Kawasima. This potential is dependent on the time-averaged kinetic and potential energies of the acoustic fields. Later, Eller \cite{eller1968force} reported theoretical analysis and experimental results on trapping the air bubble in the liquid medium under the acoustic standing wave against the upward buoyancy force. Furthermore, Crum \cite{Crum1971Jul} conducted a comprehensive theoretical and experimental investigation of the acoustic force acting on small liquid droplets (paraldehyde, hexane, benzene, toluene, chlorobenzene, and carbon tetrachloride) introduced in water under the influence of a standing acoustic wave. Crum successfully suspended the small liquid droplets using acoustic force against gravity and established that the minimum pressure amplitude required to suspend the droplet is independent of the droplet size. Coakley et al. \cite{coakley1989cell} demonstrated cell manipulation techniques and conducted theoretical analysis on the effects of acoustic pressure on the suspension position of the cells. Following the above fundamental works of acoustic radiation force (ARF) acting on particles, droplets, and cells, several practical applications have been demonstrated in the past two decades in various fields such as biological \cite{Petersson2007Jul,Wiklund2012May,Lee2015Mar,Ahmed2016Mar}, medical \cite{Augustsson2012Sep,Li2015Apr,Song2014Mar,Caplin2015Jul}, food\cite{Grenvall2009Aug,Juliano2011Sep} and chemical sciences \cite{Li2019Jan,Chen2021Nov,Wang2009Aug}. Recently, by extending Crum's work, Luo et al.\cite{luo2016experimental,luo2017suspension} experimentally investigated the effects of droplet size, acoustic pressure, frequency, and density ratio on the suspension characteristics of droplets. 

Despite the extensive literature on acoustic radiation force (ARF) experienced by the particles (beads/cells/droplets), existing works including those mentioned above are restricted to the Rayleigh limit ($a<<\lambda$, i.e. particle size '$a$' much smaller than the wavelength '$\lambda$'). Thus, the behavior of particles above the Rayleigh limit remains largely unexplored. Following are the recent works, that address the acoustic radiation force acting on the large particles. Baasch et al.\cite{Baasch2018Jan}  theoretically investigated the acoustic radiation force acting on larger particles and droplets of size up to ${a}/{\lambda }\leq 0.35$. Comparing their numerical results with ARF for small particles obtained from the Gor'kov potential \cite{gor1962forces}, they showed that the ARF equation overestimates the force as the size of the droplets or particles departs from the Rayleigh limit. Ospina et al. \cite{ospina2022particle} investigated the acoustic levitation of polystyrene particles in air using a symmetric concentric levitator and they experimentally found that the smaller particles, less than half the wavelength, are trapped along the axis around pressure nodes, while larger particles are trapped nearer to pressure antinodes. 

In this work, we experimentally investigate the suspension behavior of water droplets in an oil medium subjected to standing acoustic waves, where the droplet size is of the same order as the wavelength ($a\sim\lambda$). we particularly focus on the regime where the dynamics of the droplet is only governed by the interplay between acoustic force and the gravity force. For the droplet size beyond the Rayleigh limit, we show that the critical acoustic power ($P_{cr}$) required to suspend the droplet increases with the droplet size. This is in contrast to the case of droplets within the Rayleigh limit where the critical acoustic power required is independent of the droplet size. Furthermore, the average velocity and settling time of the droplet are also experimentally investigated by varying the acoustic power up to $P_{cr}$. Finally, we demonstrate the novel sorting method for droplets based on critical power. Our study provides new insights into the suspension characteristics of droplets beyond the Rayleigh limit and can open up new avenues for the development of sophisticated 
droplet sorting methods using acoustic fields.
\begin{figure}[h!]
  \center    \includegraphics[width=1\linewidth]{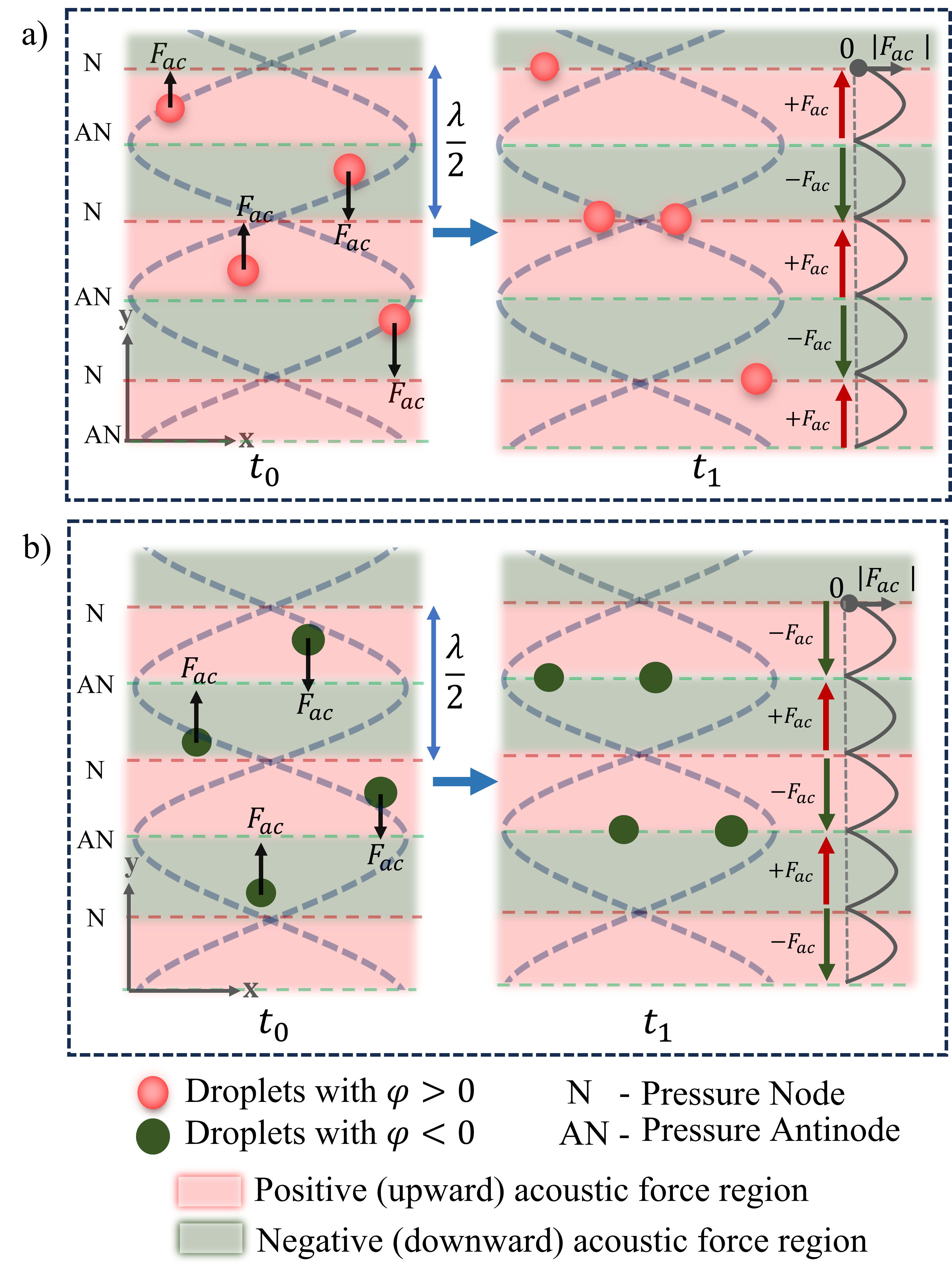} 
    \caption{ Migration of particles in the standing acoustic field given by Eq. \ref{Eq 1}. a) $F_{ac}$ acting on the positive contrast particles placed at different locations causes them to move towards the nearest node. b) $F_{ac}$ acting on the negative contrast particles placed at different locations causes them to move towards the nearest antinode. The dotted line indicates pressure variation in the $y$ direction.}
\label{phy}
\end{figure}

\section{{\textbf{Physics of the problem}}}

When a particle/droplet of size $a\ll \lambda $ is subjected to a standing acoustic wave along the $y$-direction as shown in the Fig. \ref{phy}, the primary radiation force ($F_{ac}$) acting on the particle/droplet is expressed as\cite{pangu2007droplet,pangu2004acoustically,miles1995principles,bruus2012acoustofluidics}\\
\begin{subequations}
\label{Eq 1}
\begin{equation}
\label{Eq 1a}
    F_{ac}=4\pi ka^3\varphi E_{ac}\mathrm{sin}\mathrm{}(2ky)   
,\end{equation}
\begin{equation}
\label{Eq 1b}
    \varphi =\frac{1}{3}\left(\frac{5\ \widetilde{\rho }-2}{2\widetilde{\rho}+1}-\frac{1}{\ \widetilde{\rho }\ \widetilde{c}^2}\right)  
,\end{equation}
\begin{equation}
\label{Eq 1c}
    E_{ac}={p_{a}^2}/{4{\rho }_0c^2_0}
.\end{equation}
\end{subequations}
\noindent Where $a$ is the droplet radius, $k={2\pi }/{\lambda }$  is the wave number, $\lambda $ is the wavelength of the wave, $y$ is the position of the droplet relative to the pressure node, $\varphi $ is the acoustic contrast factor, $E_{ac}$ is the acoustic energy density, $p_{a}$ is the acoustic pressure amplitude, $\widetilde{\rho}$ is the ratio of the density of the particle (${\rho}_p$) to the density of the continuous medium (${\rho}_0$), and $\widetilde{c}$ is the ratio of the speed of sound of particle ($c_p$) to the speed of sound of the continuous medium ($c_0$). When $\varphi <0$, the droplet moves toward the pressure antinode; when $\varphi >0$, it moves to the pressure node. Hereafter we refer to the pressure node and pressure antinode as simply node (N) and antinode (AN). The magnitude of the acoustic force is zero at nodes and anti-nodes within the acoustic field, while it reaches its maximum at the midpoint between the node and anti-node as illustrated in Fig. \ref{phy}. The nature of the Eqs. (\ref{Eq 1}) is clearly illustrated in Fig. \ref{phy}.  
Along with the acoustic force, gravity also determines the dynamics of the droplet. Thus, the net effect of the gravity and buoyancy acting on a droplet is given by,
\begin{equation}
\label{Eq 2}
    \ F_g=-\frac{4}{3}\pi {({\rho }_w-{\rho }_0)a}^3g\ 
.\end{equation}
Where $F_g$ is the net gravity and g is the gravitational acceleration along the negative $y$ direction. Unlike the gravity force given in Eq. (\ref{Eq 2}) (which is uniform and always acts downward along the $y$ direction), the acoustic force (Eqs. (\ref{Eq 1})) is non-uniform and its direction depends on the position of the particle. If the droplet moves in the continuous medium, it experiences the opposing drag force $F_{\mu}$ given by Hadamard-Rybczynski,\cite{lamb1924hydrodynamics}
\begin{equation}   
\label{Eq 3}
    \ F_\mu=-4 \pi\left(\frac{1+3 \tilde{\mu} / 2}{1+\tilde{\mu}}\right) \mu_{\mathrm{o}} a V \ 
,\end{equation}
where $\tilde{\mu}$ is the ratio of viscosity of the droplet ($\mu_{\mathrm{w}}$) to the viscosity of the continuous phase ($\mu_{\mathrm{o}}$) and $V$ is the velocity of the droplet.
Under the assumption that the inertial force is negligible, the balance between the different forces described above can be expressed as follows:
\begin{equation}
\label{Eq 4}
     F_{g}+F_{ac}+F_{\mu}=0
.\end{equation}
 In the presence of both gravity and an acoustic field, the dynamics of droplets are governed by the interplay between these forces, resulting in either the suspension or settling of the droplets within the medium. The velocity of the settling droplet under the influence of these forces can be calculated from Eq. (\ref{Eq 4}). The velocity becomes zero when the droplet is suspended in the continuous medium, indicating a balance between the gravity force ($F_g$) and the acoustic force ($F_{ac}$). This can be mathematically represented by substituting $F_{\mu}=0$ ($V=0$) in the Eq. (\ref{Eq 4}),
\begin{equation}
\label{Eq 5}
     F_{g}+F_{ac}=0
.\end{equation}
The above equation clearly shows that droplet/particle suspends only in the positive (upward) acoustic force region in Fig. \ref{phy}. By substituting Eqs. (\ref{Eq 1}) and (\ref{Eq 2}) into Eq. (\ref{Eq 5}), the acoustic pressure amplitude ($p_a$) required for droplet suspension can be obtained. The minimum pressure amplitude ($p_{min}$) necessary for suspension occurs at the position where the upward acoustic force is maximum (i.e. ${sin}\mathrm{}(2ky)=1$), 
\begin{equation}
\label{Eq 6}
    \ p_{\mathrm{min}}=\sqrt{\frac{2 \lambda \rho_{\mathrm{o}} c_{\mathrm{o}}^2\left(\rho_{\mathrm{w}}-\rho_{\mathrm{o}}\right) g}{3 \pi \varphi}}\  
.\end{equation}
 The relationships between acoustic power ($P_{ac}$), acoustic energy density ($E_{ac}$), and pressure amplitude ($p_{a}$) can be expressed as \cite{muller2012handbook,bruus2012acoustofluidics,Mettu2020Apr}
\begin{equation}
\label{Eq 7}
      \ P_{ac} \propto E_{ac}\propto p_{a}^{2}\
.\end{equation}
From the Eq. (\ref{Eq 6}) and Eq. (\ref{Eq 7}), the following result can be inferred,
\begin{equation}
\label{Eq 8}
     \ P_{min}\propto E_{min}\propto p_{min}^{2}\neq f(a)\
.\end{equation}
 From the above equations, it is evident that the minimum acoustic power ($P_{min}$) or minimum acoustic energy density ($E_{min}$) or minimum pressure amplitude ($p_{min}$) required to suspend the droplet is independent of the droplet size. This is because any change in droplet size scales both the acoustic force ($F_{ac}$) and the force of gravity ($F_{g}$) in the same proportion ($a^3$). It is important to note that the above formula and discussions are valid only if the droplet size is within the Rayleigh limit \textit{ }$\left(a<<\lambda \right)$. Now, we proceed to investigate the behavior of droplets beyond the Rayleigh limit under the acoustic fields and gravity.
\begin{figure*}
  \center
    \includegraphics[width=0.85\linewidth]{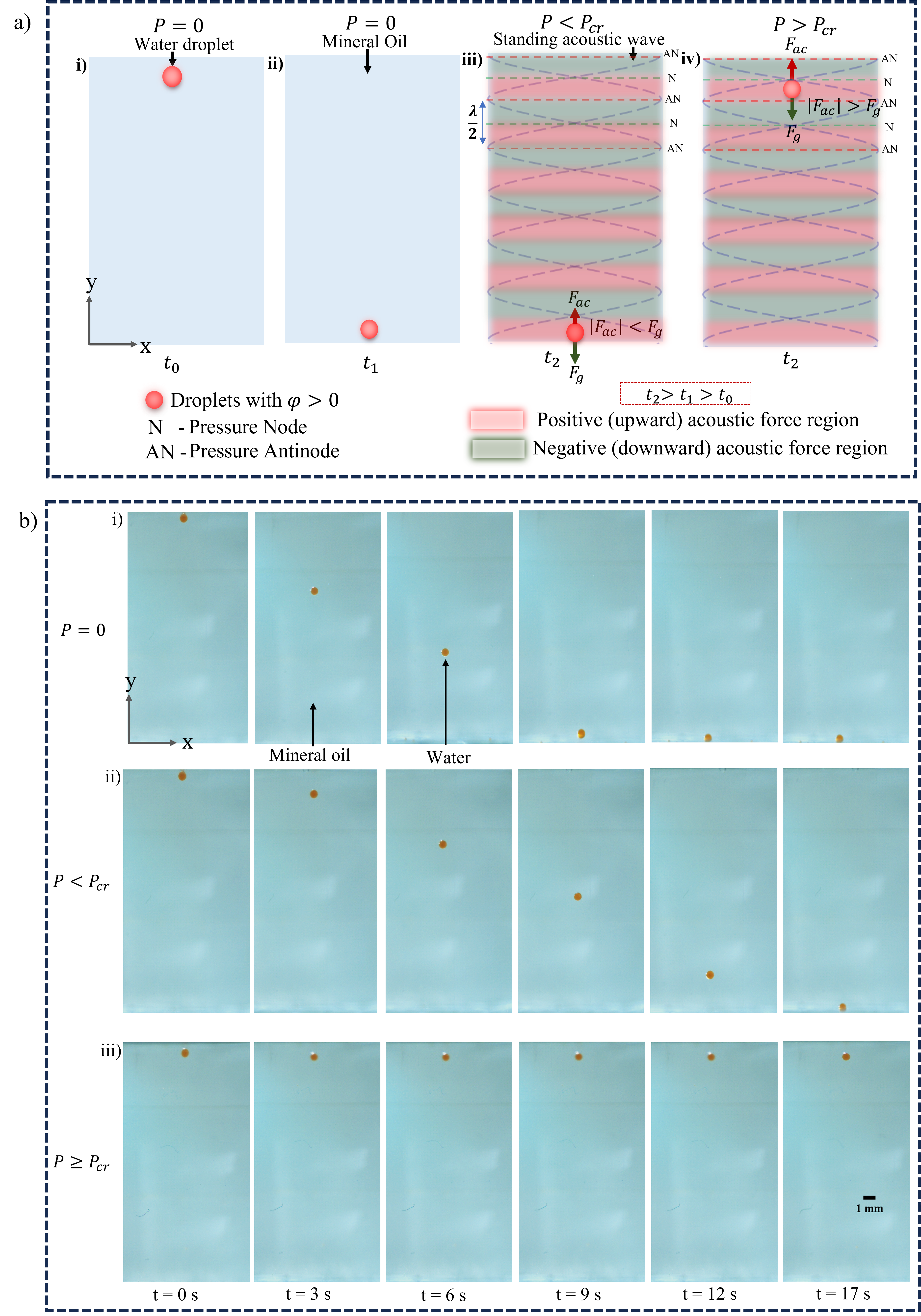} 
    \caption{Suspension characteristics of the identical size of droplets subjected to varying input power. a) Schematic representation of droplets with and without acoustic fields. b) Experimental results of suspension of the droplets i) without acoustics, ii) when subjected to input power of $1.2$ W which is less than the $P_{cr}$, and iii) when subjected to more than the critical input power of $1.5$ W.}
\label{exp}
\end{figure*}
\section{\textbf{Materials and methods}}\label{sec2}
In this study, mineral oil  (SRL chemical, India) (${\rho}_p=857.5$ $kg/m^3$, $c_p=1440$ $m/s$, and $\mu_{\mathrm{p}}=26.5$ $mPas$) is employed as the continuous medium and Dyed DI water (${\rho}_o=1000$ $kg/m^3$, $c_o=1481$ $m/s$, and $\mu_{\mathrm{o}}=1$ $mPa.s$) as the droplet/dispersed medium. Experiments are performed by introducing the above fluids in a quartz rectangular channel of a cross-section: $8$ mm width and $6$ mm breadth. The height of the channel ($H$) along the $y$ direction is $20$ mm, the bottom of the channel is sealed with a piezoelectric transducer PZT SP-4 (Sparkler Ceramics, India) using Epoxy glue, and the top is opened to the atmosphere. The outer surface of the quartz glass channel is coated with Polydimethylsiloxane (PDMS) to improve its optical clarity.  The transducer is actuated to introduce acoustic fields in the fluid domain by means of electrical excitation provided by a  power amplifier (AR RF/Microwave Instrumentation 50U1000) and a function generator (Tektronix AFG1022). The power input ($P$) to the transducer is obtained from the voltage and current measurement using the Digital Storage Oscilloscope (Tektronix TDS 2024C) and current probe (Yokogawa 701933) respectively. The experiments are conducted at the frequency ($f$) of $720\ kHz$. To capture the motion of droplets, a high-speed camera (Phantom VEO 746, USA) is used along with an LED light source (Phantom, USA) for illumination. For each test, mineral oil is initially introduced into the quartz glass channel followed by applying an acoustic wave field, and subsequently, a water droplet is introduced using the syringe for the study. The uncertainty of the distance measurement by the high-speed camera is $\pm$ $0.031$ mm (2 pixels). The input power ($P$) supplied to the square transducer of $25$ mm X $25$ mm is partially transferred ($P\propto P_{ac}$) to the fluid domain ($8$ mm X $6$ mm) in contact as the acoustic power ($P_{ac}$) while the remaining input power is transferred to the glass.

\section{\textbf{Results and discussion}}\label{sec3}
This section provides a comprehensive investigation of the trapping/suspension characteristics of aqueous droplets in an oil medium. The primary objective is to investigate how these droplets behave as their size approaches the order of the wavelength of the standing wave (beyond the Rayleigh limit).

\subsection{\textbf{ Interplay between gravity and acoustic fields}}\label{interplay}

\begin{figure}[h!]
  \center
    \includegraphics[width=0.99\linewidth]{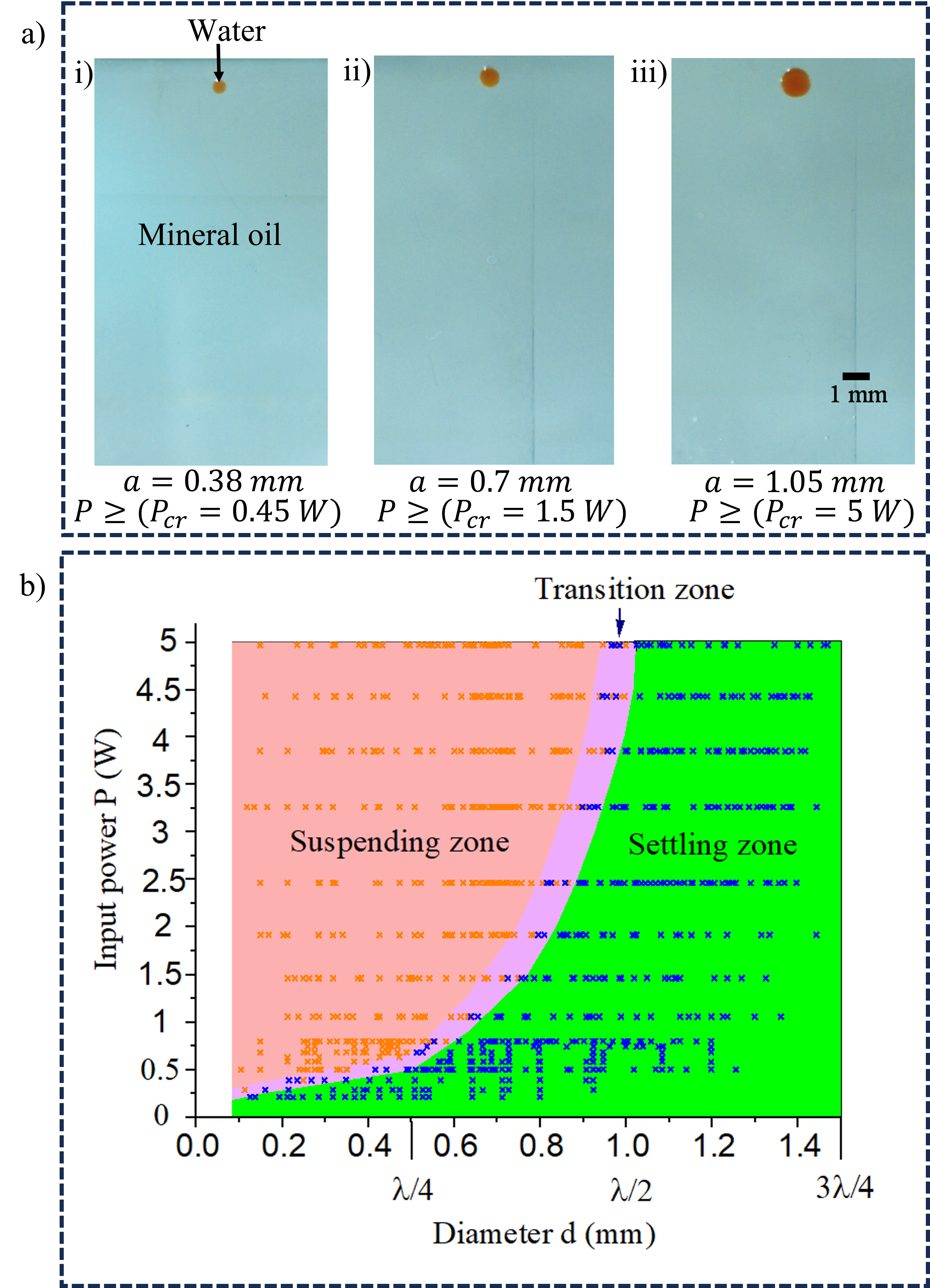} 
    \caption{ Suspending the droplet beyond the Rayleigh limit. a) Experimental results of critical input power required to suspend the droplets of different sizes, i)  $a=0.38 mm$, ii) $a=0.7 mm$, and iii) $a=1.05 mm$. b) Characterization of suspending, transition, and settling zones. The critical input power curve separates the settling and suspending zones.}
\label{cri}
\end{figure}

The dynamics of the droplet is governed by the interplay between acoustic and gravity forces. The role of the interfacial tension force is neglected since both the gravity and acoustic forces applied are not enough to deform the droplets as observed in the experiments (Figs. \ref{exp} \& \ref{cri}). Figure \ref{exp}a illustrates the droplet behavior under the influence of gravity and acoustic fields while Fig. \ref{exp}b displays the experimental results of a droplet of a specific size exposed to the acoustic field of varying power. In the absence of an acoustic field (Fig. \ref{exp}b.i), a higher-density droplet (water) in a lower-density medium (mineral oil) undergoes a uniform downward motion due to the balance between gravity and drag forces. Whereas, it is observed that the addition of an acoustic field disrupts the droplet's uniform motion. If the applied input power is strong enough to overcome the gravity, the droplet suspends (Fig. \ref{exp}a.iv and Fig. \ref{exp}b.iii). When the applied input power is insufficient, the droplet settles at a delayed time (Fig. \ref{exp}a.iii and Fig. \ref{exp}b.ii) compared to the settling time of the droplet in the absence of acoustic fields (Fig. \ref{exp}a.ii and Fig. \ref{exp}b.i). The aforementioned results are clearly explained below.
\begin{figure*}[htbp]
\center
\includegraphics[width=0.75\linewidth]{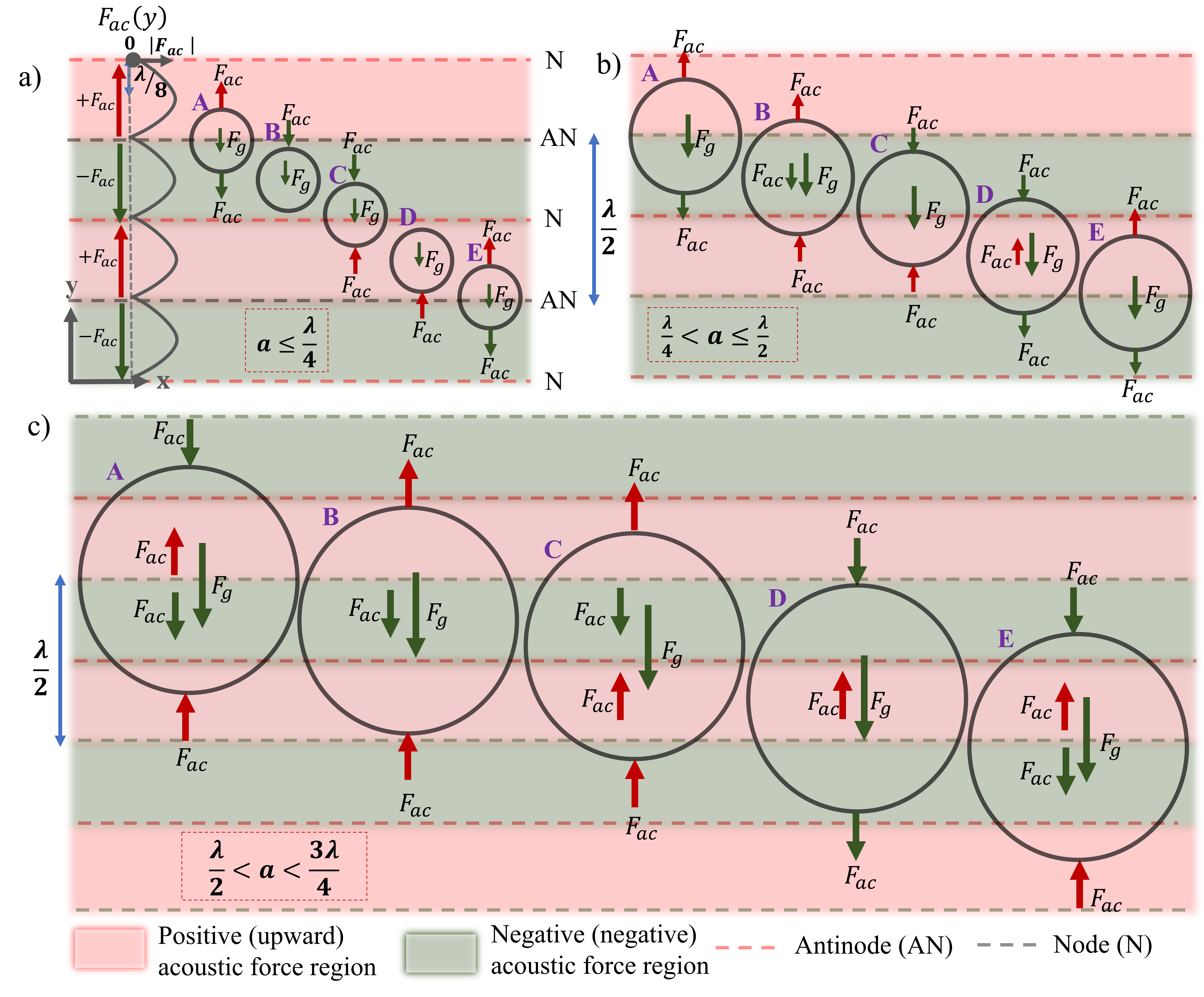} 
\caption {Schematic representation of the interplay between the acoustic and gravity forces at different droplet positions. a) $a< \frac{\lambda}{4}$, b) $\frac{\lambda}{4} < d < \frac{\lambda}{2}$, and c) $\frac{\lambda}{2} < d < \frac{\\3\lambda}{4}$.}
\label{vol}
\end{figure*}

\begin{figure*}[htbp]
\center
\includegraphics[width=1\linewidth]{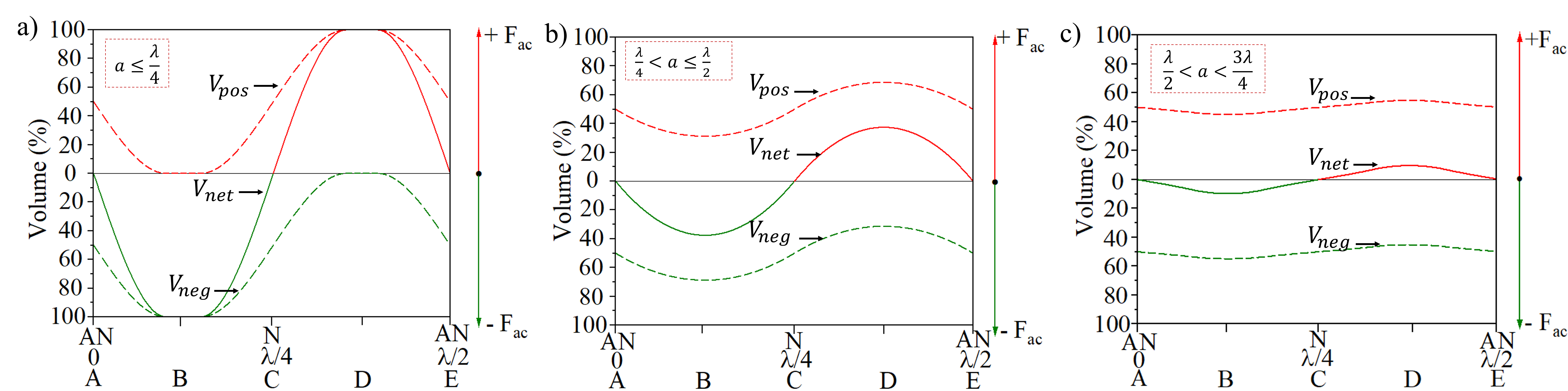} 
\caption {Analysis of net volume responsible for acoustic force for a different droplet sizes, a) $a<\frac{\lambda}{4}$, b) $\frac{\lambda}{4}<d<\frac{\lambda}{2}$, and c) $\frac{\lambda}{2}<d<\frac{\\3\lambda}{4}$. Note: The formula used to calculate the volume of the portion of the sphere is given by: volume $= \pi h^2 (3R - h)/3$ \cite{polyanin2006handbook}, where $h$ represents the height of the cap and $R$ denotes the radius of the sphere.}
\label{volg}
\end{figure*}

Since the wavelength ($\lambda=c_{0}/f=2.057$ mm) of the acoustic fields is much lesser than the height ($H=20$ mm) of the domain $\lambda\ll H $, it produces a series of nodes and anti-nodes in the fluid domain (Fig. \ref{exp}a.iii and Fig. \ref{exp}a.iv). This results in two alternating force regions, one is a positive acoustic force region where the acoustic force acts upward (the region below the node and above its nearest anti-node, as depicted in red in Fig. \ref{exp}a.iii and Fig. \ref{exp}a.iv) and the other is negative acoustic force region where the acoustic force acts downward (above the node and below its nearest anti-node, as depicted in green in the Fig. \ref{exp}a.iii). From Fig. \ref{exp}a, it is clear that when a water droplet is placed in the negative acoustic region, it will be pushed to the node since the acoustic force and gravity force acts in the downward direction. Once it comes below the node or positive acoustic region, the acoustic force starts acting upward direction opposing the gravity force. At this position, if the applied force is not enough to overcome gravity, then the droplet settles by passing through the series of nodes and anti-nodes (Fig. \ref{exp}b.ii). In this settling process, the downward velocity of the droplet becomes non-uniform as the droplet velocity is more in the negative acoustic force region and less in the positive acoustic force region. Consequently, the droplet spends more time in the positive acoustic force region and less time in the negative acoustic force region which results in delayed settling time compared to the settling time of the droplet in the absence of an acoustic field. The above settling time delay in the presence of an acoustic field is explained more clearly as follows: let's assume the average acoustic force magnitude ($\left |  F_{ac}\right |=|F|/2$) acting on the droplet is half of the gravity force ($F_g=-F$), the net force ($F_g+F_{ac}$) acting on the negative and positive acoustic force region become -$3/2F$ and -$1/2F$. If -$V$, -$l$, and $t$ are the downward velocity, downward displacement, and settling time of the droplet in the absence of acoustic force, then the velocity in the negative and positive acoustic regions are -$3/2V$ and -$1/2V$ respectively. Thus, the time the droplet spends on the negative acoustic region (-$l/2$) and positive acoustic region (-$l/2$) are $t/3$ and $t$, the total time taken by the droplet for settling becomes $4t/3$ as compared to the time $t$ taken by the droplet in the absence of acoustic fields.It is observed that as the power increases, settling time increases when $F_{ac}$ approaches the dominant $F_g$.
When the applied input power ($P\geq P_{cr}$) generates sufficient acoustic force to overcome gravity, the droplet suspends in the positive acoustic force region (Fig. \ref{exp}a.iv). 

The above discussion is clearly evident in the experimental results shown in Fig. \ref{exp}b, the droplet with a size of $0.7$ mm is introduced into the channel at time $t=0$ s. In the absence of acoustic fields (Fig. \ref{exp}b.i), the droplet settles at $9$ s due to the balance between gravity and drag forces. When an input power of $1.2$ W is applied which is below the critical power ($P<P_{cr}$) (Fig. \ref{exp}b.ii), the motion of droplets in a series of nodes and antinodes results in a delayed settling time of $17$ s. When the input power is more than the critical input power of $1.5$ W ($P\ge P_{cr}$) for the $0.7$ mm droplet size, the acoustic force becomes sufficiently strong to overcome gravity, leading to the suspension of the droplet at the positive acoustic force region.
\subsection{\textbf{ Beyond the Rayleigh limit}}\label{sec2p3}
Figure \ref{cri} shows the experimental results on droplets of different sizes (beyond the Rayleigh limit) subjected to the acoustics fields. The results are remarkable, as the droplet size increases, the critical power $P_{cr}$ required to suspend the droplet increases exponentially. This is in contrast to Eqs. (\ref{Eq 6}) and (\ref{Eq 8}) which predicts that the minimum power required to suspend the droplet is not the function of the droplet size.

In Fig. \ref{cri}b, the relationship between droplet size and critical acoustic power demonstrates distinct trends across different size ranges. For droplet sizes up to a quarter of the wavelength $\left(a<{\lambda}/{4}\right)\ $, a slight increase in the critical acoustic power is observed as shown in Fig. \ref{cri}b. Whereas, for droplet sizes larger than a quarter of the wavelength $\left(a>{\lambda}/{4}\right)\ $, the critical acoustic power follows an exponential-like curve. For example, a droplet size of $0.6$ mm diameter requires $0.8$ W of critical acoustic power, and a droplet size of $0.9$ mm diameter demands critical power of $4$ W. When attempting to suspend droplets with a size $a>\lambda/2$, the input power required increases quite significantly. However, at $5$ W, cavitation occurred and limited further power increment in our experiment.


 The behavior of the droplets beyond the Rayleigh limit observed in Fig. \ref{cri}b can be explained qualitatively to a large extent by assuming the bigger droplet is a collection of Rayleigh particles/droplets. By adopting this assumption, we can apply the small particle acoustic field to every point of this larger droplet (Fig. \ref{phy}). The immediate consequences of the assumption are as follows: the force acting on the smaller droplet at the node is zero by Eqs. (\ref{Eq 1}). Whereas, for the larger droplet placed at the node, the acoustic force acting on the upper portion of the droplet is negative and the lower portion of the droplet is positive as shown in Fig. \ref{vol}a (position A). Thus, the net acoustic force acting on the droplet becomes zero. By using the above assumption, first, we proceed to explain the migration of the larger droplets $a<\lambda/4$ and followed by $a>\lambda/4$.

 Let's assume the center of the droplet size of $a<\lambda/4$ is initially introduced at the AN as shown in Fig. \ref{vol}a (position A). The net $F_{ac}$ acting on the droplet at position A is zero as +$F_{ac}$ experienced by the upper portion is counterbalanced by the -$F_{ac}$ of the bottom portion. Thus, the dominant gravity force moves the droplet downwards. At position B, the droplet moves downwards with increased velocity as the acoustic force on the entire droplet acts downwards supporting gravity. At position C also droplet continues to move downwards, as the force balance scenario is similar to Position A. At position D, as the acoustic force acting opposite to the direction of gravity force, the droplet can be suspended, if the applied acoustic power  $\ge P_{cr}$. The reason for the increase in the $P_{cr}$ along with the droplet size can be explained as follows: First for droplets $a<\lambda/4$, as the droplet becomes bigger the droplet volume is distributed in the lesser force magnitude region as shown in the Fig.\ref{vol}a. For a given power, the average acoustic force acting on the bigger droplet is less compared to the smaller droplet as shown in Fig. \ref{vol}a. Thus $P_{cr}$ for a bigger droplet will be slightly more compared to the smaller droplet. This explains the marginal increases in $P_{cr}$ for the droplet size $a<\frac{\lambda}{4}$ as observed in Fig.\ref{cri}. For the case of $a<\lambda/4$, the droplet can be completely accommodated in the positive acoustic force region as shown in Fig. \ref{vol}a (at position D).
 
 Similar to the droplet size $a<\lambda/4$, droplets of the size of $a>\lambda/4$ are also suspended in Position D shown in Fig. \ref{vol}b \& Fig. \ref{vol}c where the major portion of the droplet is present in the positive acoustic force region. However in the case of the droplet size  $a>\lambda/4$, the droplet can't be completely accommodated in a positive acoustic force region, some portion of the droplet is always present in the negative acoustic force region. Thus to suspend the droplet size of $a>\lambda/4$, the positive acoustic force acting on a portion(s) of the droplet not only opposes the gravity but must also counteract the negative acoustic force acting on the other portion(s) of the droplet as shown in Fig. \ref{vol}b \& c. Because of this reason, $P_{cr}$ rises exponentially when the droplet size increases more than $\lambda/4$. For $a>\lambda/4$, the net volume responsible for the resultant acoustic force is significantly less than the total volume of the droplet which can be explained by the volume distribution analysis given below. 

 The volume distribution analysis of different droplet size ranges $a<\frac{\lambda}{4}$, $\frac{\lambda}{4} < d < \frac{\lambda}{2}$, and $\frac{\lambda}{2} < d < \frac{\\3\lambda}{4}$ are shown in the Fig. \ref{volg}. The volume analysis presented here is approximate since the acoustic force magnitude is assumed to be uniform. The net volume of the droplet ($V_{net}$) responsible for the acoustic force is $V_{net}=\left | V_{pos}-V_{neg} \right |$, where $V_{pos}$ is the volume portion of the droplet in the positive acoustic force region and $V_{neg}$ is the volume portion of the droplet in the negative acoustic force region. The variation of the $V_{net}$ and the direction of force acting on it with respect to the position of the droplet is shown (solid line) in Fig. \ref{volg}. The variation is shown for a ${\lambda}/{2}$ cycle (from one AN to consecutive AN) and the same pattern repeats throughout the domain. Red and green dotted lines indicate the percentage of droplet volume in the positive and negative acoustic force region respectively. From Fig. \ref{volg}a it is clear that for $a\leq\lambda/4$, the whole droplet ($V_{net}=100\%$) volume experiences the positive acoustic force at the position D. Whereas for droplets with sizes between $\frac{\lambda}{4} < d\leq \frac{\lambda}{2}$, the maximum $V_{net}$ is significantly less than $100\%$. For instance, if the droplet size is $a=\lambda/2$, the maximum $V_{net}$ experiencing the upward acoustic force is $37.5\%$ of the total volume at position D (Fig. \ref{volg}b). Similarly, when the droplet size between $\frac{\lambda}{2} < d < \frac{\\3\lambda}{4}$, the $V_{net}$ is even smaller compared to the previous case. For example, if the droplet size is $a=\frac{\\13\lambda}{20}$, the maximum $V_{net}$ experiencing the upward acoustic force is $9.69\%$ of the total volume at position D (Fig. \ref{volg}c). From this analysis, it is clear that as the droplet size increases acoustic force becomes ineffective (requires more power) in suspending the droplet against gravity. 
 
\subsection{\textbf{Settling time and average velocity}}\label{velocity}

\begin{figure}[h!]
  \center
    \includegraphics[width=1\linewidth]{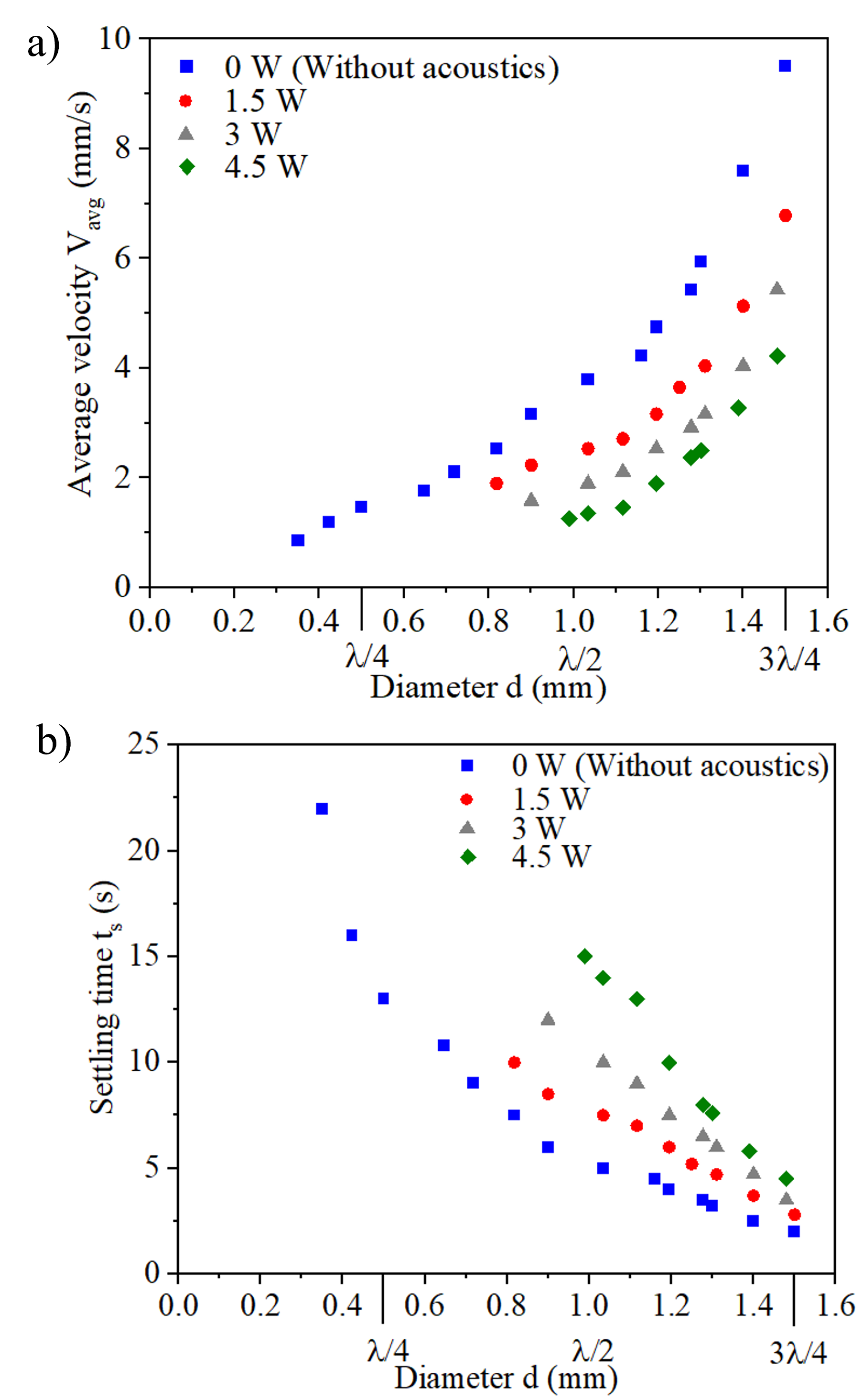} 
    \caption{ Experimental results of settling time (a) and average velocity (b) of the droplets of different sizes subjected to varying input power. Note: Applied input power is less than critical input power for the above experimental results.}
\label{vel}
\end{figure}

In this section, we study the average velocity ($V_{avg}$) and settling time ($t_s$) of the droplet when the input power applied is less than $P_{cr}$. From Fig. \ref{vel}a, it is evident that the presence of acoustic fields delays the setting time of the droplet, as the input power increases the settling time increases. The reason for the delayed settling time is clearly explained in Section \ref{interplay}. From Fig. \ref{vel}a, it can be inferred that the influence of input power on the settling time gets weaker as the droplet size increases since the droplet covers multiple positive and negative acoustic force regions. The average velocity shown in Fig. \ref{vel}b is in line with the settling time results. The acoustic field reduces the average velocity of the droplet as compared to the uniform velocity of the droplet due to gravity without acoustic fields. This is attributed to the fact that the droplet spends more time in the negative acoustic force region compared to the positive acoustic force region. It is also to be noted that as the droplet size increases, the settling time decreases and average velocity increases for any input power including the zero power (without acoustics).

\subsection{\textbf{Sorting of the droplets based on the critical acoustic power method}}\label{sorting}

\begin{figure}[h!]
  \center
    \includegraphics[width=0.99\linewidth]{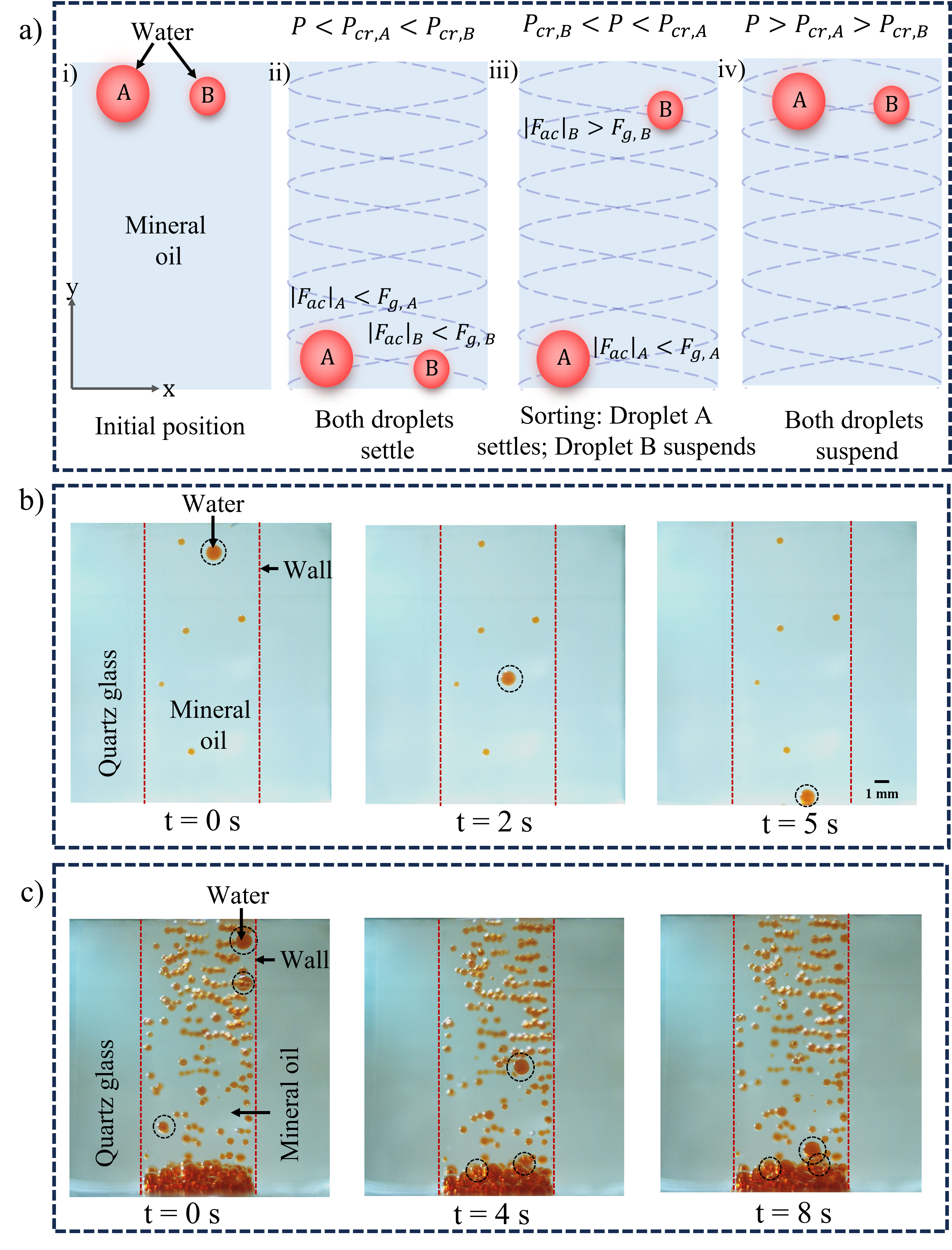} 
    \caption{Droplet sorting based on the critical power method. a) Schematic representation. Experimental result of droplet sorting: b) in a diluted w/o emulsion and c)  in a dense w/o emulsion.}
\label{sort}
\end{figure}
If the size of the droplets is beyond the Rayleigh limit, it offers a novel sorting method called sorting based on critical power. Before proceeding to understand this method, it is crucial to understand the two conventional sorting methods widely used for small particles based on Eqs. (\ref{Eq 1}): Size-based sorting and contrast factor-based sorting. In contrast factor-based sorting \cite{petersson2005continuous}, acoustic radiation force is utilized to direct particles of positive contrast factor to node and negative contrast factor to anti-node. On the other hand, in size-based sorting \cite{petersson2007free}, since ARF given in Eqs. (\ref{Eq 1}) is proportional to $a^3$, the larger particles tend to move faster towards the node or anti-node compared to the smaller particles. Size-based sorting is achieved by the timely collection of large particle from the node or antinode before smaller particle reaches there. It is important to note that if enough time is given, both the smaller and larger particles will eventually converge towards the node or anti-node.

 The proposed sorting method based on critical power is illustrated in Fig. \ref{sort}a. Both the droplets settle if the applied power ($P$) is less than the critical power of both the droplets ($P<P_{cr,B}<P_{cr,A}$) (Fig. \ref{sort}a.i). Droplet B suspends and droplet A settles (Fig. \ref{sort}a.iii) if $P_{cr,B}<P<P_{cr,A}$. In the case of $P>P_{cr,A}>P_{cr,B}$, both droplets suspend as shown in Fig. \ref{sort}a.iv. Here we propose a sorting method based on the condition illustrated in Fig. \ref{sort}a.iii). The above condition can also be stated as follows: For the given input power, there exists a critical diameter ($d_{cr}$), whereby droplets with a diameter less than the $d_{cr}$ suspend and droplets with size more than the $d_{cr}$ settle (Fig. \ref{sort}a.iii).

Fig. \ref{sort}b \& Fig. \ref{sort}c experimentally demonstrate sorting in diluted w/o emulsion and dense w/o emulsion respectively. In Fig. \ref{sort}b, sorting occurs at an acoustic power of $1.5$ W which corresponds to a critical diameter of $0.7$ mm, and droplets of size smaller than $a < 0.7$ mm are suspended, while droplets of size larger than $a > 0.7$ mm (marked in a dotted circle) are settled within a few seconds. Similarly, in Fig. \ref{sort}c (dense w/o emulsion) the droplets of size less than $0.8$ mm and droplets of size greater than $0.8$ mm (marked in a dotted circle) are sorted by applying an acoustic power of $2.5$ W ($d_{cr}=0.8$ mm). Remarkably, this sorting method is robust and works even in dense suspension where the particle-particle interaction is significant.

\section{\textbf{Conclusion}}\label{sec4}    
In this work, we experimentally investigated the suspension behavior of droplets beyond the Rayleigh limit when subjected to standing acoustic waves. We showed that if the droplet size exceeds the Rayleigh limit, the critical acoustic power required to suspend the droplets against gravity strongly depends on the droplet size. The suspension characteristics of the droplet under different regimes were explained qualitatively by adopting the assumption that the larger droplet can be considered as a collection of Rayleigh particles/droplets. In addition, we also demonstrated the novel sorting of droplets using the critical power method. Our study provides new insights into the suspension characteristics of the droplets beyond the Rayleigh limit which will aid the development of advanced droplet sorting techniques using acoustic fields. To enhance our understanding of the current experimental results and explore droplet behavior beyond our experimental constraints, we are conducting numerical simulations, to be discussed in our forthcoming research.

\begin{acknowledgments}
This work is supported by the Department of Science \& Technology - Science and Engineering Research Board (DST-SERB) via Grant No: SRG/2021/002180 and the Department of Science \& Technology - Fund for Improvement of Science \& Technology Infrastructure (DST-FIST) via Grant No: SR/FST/ET-I/2021/815. We express our gratitude to Mr. Aswinraj M from IIITDM for his support with the electrical circuit and measurements.
\end{acknowledgments}

\section*{Author Declarations}
\subsection*{Conflict of Interest}
The authors have no conflicts of interest to disclose.

\section*{Data Availability Statement}
The data that support the findings of this study are available from the corresponding author upon reasonable request.

\section*{References}
\nocite{*}
\bibliography{pof}

\end{document}